\documentclass[9pt,twocolumn,twoside]{pnas-new-custom}
\usepackage[nameinlink,capitalize]{cleveref}
\usepackage{mhchem}
\usepackage{balance}
\usepackage{array}
\newcolumntype{L}[1]{>{\raggedright\arraybackslash}p{#1}}
\articletype{SOCIAL SCIENCES}
\templatetype{pnasresearcharticle}
\setboolean{displaywatermark}{false}

\begin{document}
\title{An index to quantify scientific debt}

\author[a,*]{N. Zen}
\affil[a]{The author's affiliated institute did not regard this work as appropriate research for the institute. For this reason, the author's affiliation is not disclosed. The institute assumes no responsibility for its content.}

\leadauthor{Zen}

\significancestatement{Scientists disseminate reliable knowledge. This achievement is quantified by the $h$-index and valued by funding agencies and hiring committees. By contrast, no index yet quantifies the dissemination of false knowledge; I propose one here. The $z$-index accumulates as misinformation spreads. In an era of widespread research misconduct, from classical data fabrication to AI-assisted paper mills, the $z$-index may become a key tool for limiting that spread. Analysis using the $z$-index also showed that considerable debt has accumulated in the two leading journals. This may not be surprising to some readers, but making it visible is nevertheless important. The analysis points to the importance of responsible data sharing.{\if0 Journals that benefit from high impact factors should also bear corresponding responsibility.\fi}}

\authorcontributions{N.Z. designed research, performed research, analyzed data, and wrote the paper.}
\authordeclaration{The author declares no competing interest.}
\correspondingauthor{\textsuperscript{*}To whom correspondence should be addressed. E-mail: {\fontfamily{DejaVuSansMono-TLF}\selectfont zen137a0@gmail.com}}


\begin{abstract}
I propose the index $z$, defined as $z=\sum\limits_{i=1}^{n} \mathrm{IF}_{i}$, where $\mathrm{IF}_{i}$ is the impact factor of the journal in the publication year of the $i$-th retracted paper. Whereas the $h$-index is a forward-looking measure of scientific achievement, the $z$-index quantifies scientific debt. The quantity $h-z$ may provide a more balanced assessment of scientific value.
\end{abstract}

\dates{This manuscript was compiled on \today}
\doi{\url{www.pnas.org/cgi/doi/10.1073/pnas.XXXXXXXXXX}}

\maketitle
\thispagestyle{firststyle}
\ifthenelse{\boolean{shortarticle}}{\ifthenelse{\boolean{singlecolumn}}{\abscontentformatted}{\abscontent}}{}

\Firstpage
The $h$-index~\cite{Hirsch2005} reflects the extent to which a researcher has produced influential publications, as measured by citation counts, throughout their career up to the present. Researchers with higher $h$-indices are generally regarded as having made greater contributions to science, and their future work is often expected to maintain a similarly high level of impact~\cite{Hirsch2007,Ball2007}. Consequently, the $h$-index is likely to be a metric of considerable interest to funding agencies and academic hiring committees~\cite{Acuna2012,Hirsch2020}.

However, despite being adopted as an official metric by scientific databases such as Google Scholar, the $h$-index is not always scored appropriately. One such case arises when papers that contribute to a researcher's $h$-index are later retracted. Retraction implies that, at a minimum, the conclusions of the paper can no longer be scientifically supported. It is therefore difficult to justify why such papers continue to contribute to the $h$-index.

An example may help illustrate the issue. The field of superconductivity, in which the author works, has experienced several cases of research misconduct. A recent case attracted considerable attention and raised concerns about the credibility of the field. The events began in 2020, and the last paper in the series was retracted in 2024. The primary basis for the retractions was the manipulation of research data, and the case ultimately concluded with responsibility being primarily attributed to Ranga Dias~\cite{vanKampen2022,Hand2022,GaristoAPS2023,vanKampen2023,NYTimes2024}. This paper is not intended to criticize or disparage any individual. Rather, it is motivated by the hope that lessons can be learned from \emph{any} unfortunate episode and applied for the benefit of the scientific community in the future.

His $h$-index is 19 (Google Scholar, 25 July 2026)~\cite{RangaH}. This score includes two papers retracted from {\it Nature}~\cite{RangaNat2020,RangaNat2023}, two from {\it Physical Review Letters}~\cite{RangaPRL2021-1,RangaPRL2021-2}, and one from {\it Chemical Communications}~\cite{RangaChemComm2022}. Simply treating these five papers as if they had never existed{\if0 and subtracting their contribution from the current $h$-index\fi} may not provide an appropriate assessment of scientific responsibility. As noted above, the $h$-index is inherently forward-looking: a high $h$-index is often interpreted as evidence that a researcher is likely to continue producing influential work in the future. By the same logic, a history of retractions should also be regarded as relevant information when evaluating scientific value. The $z$-index is proposed as the dual counterpart of the $h$-index. The $h$-index measures achievement, and the $z$-index measures debt.

Retracting a paper published in a journal with an impact factor (IF) of 1 is not equivalent to retracting a paper published in a journal with an IF of 50. Results published in high-impact journals are more likely to be widely disseminated beyond their immediate field~\cite{MagdalenaSkipper}, and their retraction can therefore have a correspondingly greater negative impact. In this context, the IF is useful because it can be regarded as an expected measure of the influence that a paper published in that journal will have~\cite{Garfield2006}. Larger expectations should correspond to larger debt. Accordingly, the $z$-index can be expressed as
\begin{equation}
z=\sum_{i=1}^{n} \mathrm{IF}_{i},\label{eq:z}
\end{equation}
\Parasplit
where $\mathrm{IF}_{i}$ denotes the impact factor of the journal in the publication year of the $i$-th retracted paper. The $h$-index and $z$-index are naturally complementary because both are citation-based metrics.

In the case of Ranga Dias, one {\it Nature} paper published in 2020~\cite{RangaNat2020}, two {\it Physical Review Letters} papers published in 2021~\cite{RangaPRL2021-1,RangaPRL2021-2}, one {\it Chemical Communications} paper published in 2022~\cite{RangaChemComm2022}, and one {\it Nature} paper published in 2023~\cite{RangaNat2023} were retracted. Using the impact factors of the corresponding journals in the years of publication~\cite{Clarivate}, the $z$-index is calculated as $z = 49.962 + 9.185 + 9.185 + 4.9 + 50.5 \approx 124$. Given his $h$-index of 19, the corresponding $h-z$ value is $19 - 124 = -105$. A negative value indicates that future work may have a net negative impact on science, and its magnitude reflects the extent of that impact.
\section*{Potential Concerns}
Three potential concerns may arise when introducing the $z$-index. Let us consider each of these issues in turn.
\subsection*{Treatment of coauthors}
The treatment of coauthors in the $z$-index follows the same principle used in the calculation of the $h$-index. That is, the same $z$-index is assigned to all coauthors, not only to the corresponding author or to the individual primarily responsible for the retraction.

The treatment of coauthors was already recognized as an issue when the $h$-index was introduced~\cite{Hirsch2005,vanRaan2006,Egghe2008,Schreiber2009,Hirsch2010hbar} and remains a topic of discussion today~\cite{Hirsch2019,Tietze2020,Tzitzikas2024}. However, the $h$-index has not been displaced by any alternative. Indeed, the $h$-index treats all coauthors equally. This is also true for papers in high energy physics and quantum computing, which often have dozens or even hundreds of coauthors. Since all coauthors equally benefit from the $h$-index, all coauthors should likewise share responsibility under the $z$-index.

This may seem unfair to coauthors who were not involved in any misconduct. However, such treatment could strengthen the reliability of science by encouraging greater scrutiny among coauthors. If coauthors know that they will receive the same $z$-index as the principal actor, they will be strongly motivated to examine one another's data and analyses more carefully. Internal reporting among coauthors is particularly important for preventing problems from developing further, given the roles and responsibilities that come with being credited as authors~\cite{ICMJE-authorship}.

\subsection*{Author resistance to retraction}
Once a paper is retracted, the corresponding $z$-index increases and the author's scientific value, as measured by $h-z$, decreases. A legitimate concern is that the introduction of the $z$-index could increase author resistance to retractions and thereby make it more difficult to remove unsupported publications from the scientific literature.

The case of Ranga Dias is instructive. He strongly opposed the scrutiny of his work, employing measures ranging from refusing to release raw data on the grounds that patent applications were pending~\cite{vanKampen2022,GaristoAPS2023} to sending cease-and-desist letters to the institutions of individuals involved in tracking the underlying data~\cite{vanKampen2022,Hand2022}. Nevertheless, the paper was ultimately retracted despite opposition from all authors~\cite{RetNoteRangaNat2020}. Once evidence of data manipulation becomes sufficiently clear~\cite{Marel-Hirsch2022,Marel-Hirsch2023}, retractions cannot be prevented{\if0, whether or not the $z$-index exists\fi}.

The example above may represent a special case in which those tracking the data were substantially more capable than the authors. What, then, if the authors are substantially more accomplished---for example, Nobel Prize laureates? The answer is the same. Even when Nobel Prize laureates are involved, retraction becomes unavoidable once evidence of data manipulation comes to light. For the two Nobel laureates Gregg Semenza ($h=223$)~\cite{SemenzaH} and Thomas Südhof ($h=238$)~\cite{SudhofH}, potential signs of data manipulation in 23 and 2 papers, respectively, have been reported by {\it For Better Science}~\cite{Schneider2020-sem,vanKampen2023-sud}, with many other papers flagged on {\it PubPeer}~\cite{PubPeer-sem,PubPeer-sud}. At present, 15 papers by Semenza~\cite{sem1,sem2,sem3,sem4,sem5,sem6,sem7,sem8,sem9,sem10,sem11,sem12,sem13,sem14,sem15} and 2 by Südhof~\cite{sud1,sud2} have been retracted. Since even Nobel Prize laureates are not exempt from retraction, it is difficult to argue that ordinary researchers could avoid retraction through resistance. This remains true regardless of whether the $z$-index is introduced.

Their current $z$-indices are 113 and 24, respectively, while their corresponding $h-z$ values are 110 and 214, both still positive owing to their high $h$-indices.

\subsection*{Treatment of ``honest error'' retractions}
Few would argue today that scientific paper retractions are driven primarily by honest error. Although limited to the biomedical and life sciences literature, a 2012 study revealed that 67.4\% of retractions were attributable to misconduct, whereas only 21.3\% resulted from honest error~\cite{Fang2012}. Whether these proportions have increased or decreased over the fourteen years since then is an interesting question, but it does not affect the calculation of the $z$-index. The $z$-index treats honest error and misconduct in the same manner; their relative proportions do not matter. The $z$-index is intended to quantify the resulting scientific debt. Even in cases of honest error, incorrect scientific information has been disseminated and may influence subsequent research. Consequently, such cases also contribute to the $z$-index.

However, it would be inappropriate to treat a paper that remained uncorrected for one year in the same way as a paper that remained uncorrected for ten years. The duration between publication and retraction must therefore be taken into account. This is addressed in the next section. In general, papers retracted because of honest error tend to be withdrawn more quickly~\cite{COPE2025}, and their corresponding $z$-indices are therefore expected to remain relatively small.
\section*{Accounting for Retraction Delays}
Retracted papers are known to continue to be cited as valid work even after retraction~\cite{Budd1998,BarIlan2018,Candal2020,Hsiao2022}. This suggests that scientific debt accumulates with time. To account for this effect, we introduce a temporal weighting factor and define a modified index,
\begin{equation}
\tilde{z}=\sum_{i=1}^{n} \mathrm{IF}_{i}\times f_{i}(t),\label{eq:zt}
\end{equation}
where $t$ denotes the time elapsed between publication and retraction, and $f_{i}(t)$ is a weighting function of the $i$-th retracted paper.{\if0 The modified index $\tilde{z}$ provides a more appropriate measure of scientific debt by accounting for the duration over which invalid findings remain in circulation.\fi}

Retraction is often a time-consuming process. In many cases, it requires not only agreement between the authors and the journal, but also the findings of institutional or independent investigations. Nevertheless, if retractions continue to take several years, as is often the case at present, effective correction of the scientific record cannot be expected. From this perspective, providing an incentive for retraction within one year of publication ($t<12$) appears to be a reasonable way to encourage early correction and minimize the accumulation of scientific debt.

On the other hand, journal evaluation has long relied on a two-year citation window, as exemplified by the calculation of impact factors~\cite{Garfield2003}. If a paper remains unretracted throughout this period, a substantial portion of its scientific influence may already have been established. From this perspective, setting $f(24)=2$ may represent a reasonable compromise.

\begin{figure}[h!]
\centering
\includegraphics[width=0.75\linewidth]{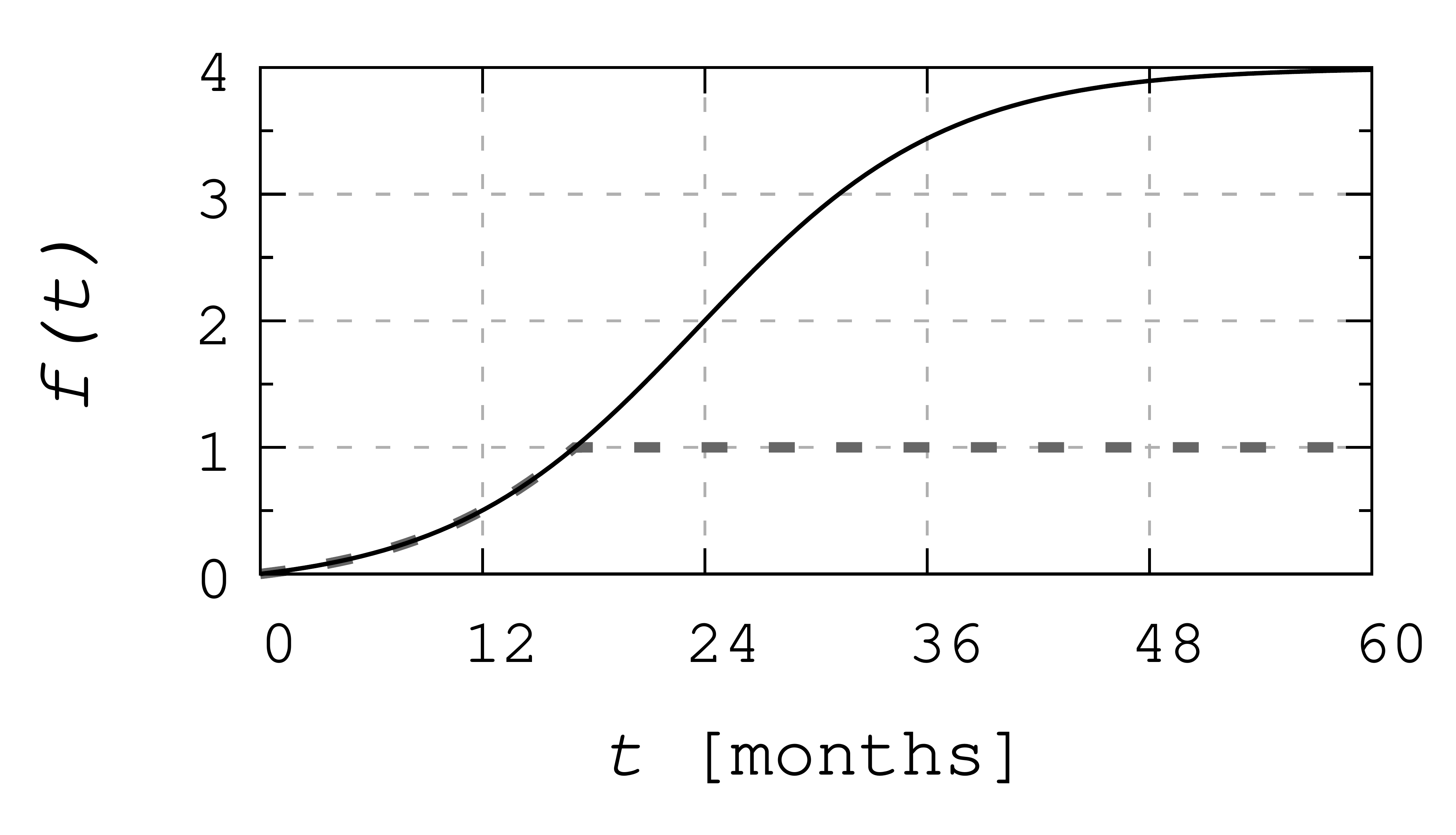}
\caption{Solid line: temporal weighting function $f(t)$. Dashed line: author-specific temporal weighting function $f_{A}(t)$. See the text for details.}
\label{fig1}
\end{figure}

The continued citation of retracted papers remains a persistent problem even today~\cite{Hsiao2022}. However, one study suggested that the influence of indirect citations beyond direct citation chains may be limited~\cite{vanderVet2016}. From this perspective, a logistic function that approaches a finite saturation value as $t \to \infty$ appears to be a reasonable choice for $f(t)$. We therefore define
\begin{equation}
f(t)=4\frac{s(t)-s(0)}{1-s(0)},\label{eq:f}
\end{equation}
where
\begin{equation}
s(t)=\frac{1}{1+e^{-0.149(t-23.6)}}.\label{eq:s}
\end{equation}
The resulting function is shown as the solid curve in \cref{fig1}. The parameters in \cref{eq:f,eq:s} were chosen such that $f(0)=0$, $f(12)=0.5$, $f(24)=2$, and $f(\infty)=4$; that is, papers retracted within one year receive a weighting factor of less than 0.5, a two-year delay results in a factor of 2, and the maximum factor is 4.

A prominent example of an honest-error retraction is that of Frances Arnold, who received the 2018 Nobel Prize in Chemistry. She attributed the error to having paid insufficient attention to the submission amid the excitement surrounding the Nobel Prize~\cite{Oransky2025}. The paper was retracted in 2020 because the reported results could not be reproduced~\cite{Cho2019}. Since the paper was published in {\it Science} in 2019, when the journal's IF was 41.846~\cite{Clarivate}, and this remains her only retracted paper, her $z$-index is 42 ($h=158$)~\cite{ArnoldH}. However, the paper was retracted only eight months after publication. Incorporating temporal weighting reduces her scientific debt substantially, yielding $\tilde{z} = 41.846 \times f(8) \approx 10$. Consequently, her $h-\tilde{z}$ value is $158-10=148$, demonstrating that a promptly corrected honest error has only a minor impact on the overall evaluation of a scientist.

Let us now recalculate the $z$-indices of the individuals mentioned above by incorporating temporal weighting. For Ranga Dias, the original $z$-index was calculated as $z = 49.962 + 9.185 + 9.185 + 4.9 + 50.5 \approx 124$. The corresponding retraction delays were 23, 39, 26, 18, and 8 months, respectively~\cite{RangaNat2020,RangaPRL2021-1,RangaPRL2021-2,RangaChemComm2022,RangaNat2023}. Applying the temporal weighting function yields $\tilde{z} = 49.962 \times f(23) + 9.185 \times f(39) + 9.185 \times f(26) + 4.9 \times f(18) + 50.5 \times f(8) \approx 165$, representing a 33\% increase in scientific debt. This case illustrates how delays in retraction can substantially increase scientific debt. His current $h-\tilde{z}$ value is $19 - 165 = -146$~\cite{RangaH}. For the two Nobel laureates, Gregg Semenza ($h=223$)~\cite{SemenzaH} and Thomas Südhof ($h=238$)~\cite{SudhofH}, the same analysis yields $\tilde{z}$ values of 450 and 62, respectively. Their corresponding $h-\tilde{z}$ values are therefore $-227$ and 176. The fact that a Nobel Prize laureate can exhibit a strongly negative $h-\tilde{z}$ value is both striking and informative.

In the examples discussed so far, $f(t)$ appears to work well. In fact, however, it has a potential drawback. Even if authors recognize an error and wish to retract the paper promptly, a journal may delay the retraction for its own reasons. Moreover, if the authors discover an honest error only ten years later, the $\tilde{z}$-index is already subject to a fourfold weight, which may make them hesitate to report the error, contrary to the intended purpose of the index. For these reasons, the dashed curve in \cref{fig1}, hereafter denoted by $f_{A}(t)$, appears to be more appropriate as the temporal weighting function for the author $\tilde{z}$-index. In this formulation, the temporal weighting factor never exceeds 1, meaning that authors can initiate the retraction process at any time without hesitation. At the same time, when $t$ is roughly less than 1.5 years, the $z$-value is discounted, thereby still encouraging early retraction.
\section*{Journal z-index}
The $h$-index is used to evaluate not only researchers but also journals. Therefore, its dual counterpart, the journal $z$-index, exists. \Cref{tab:top} presents the top 20 journals ranked by $h$-index (SCImago, 25 July 2026)~\cite{SCImago}, together with their $z$- and $\tilde{z}$-indices. A journal $h$-index of 1495 means that the journal has published 1495 papers that have each received at least 1495 citations.
\begin{table}[t]
\centering
\caption{Top 20 journals ranked by $h$-index~\cite{SCImago}}
\label{tab:top}
\begin{tabular}{lccc}
Journal & $h$ & $z$ & $\tilde{z}$\\
\midrule
Nature & 1495 & 3650 & 8290\\
Science & 1433 & 2809 & 6350\\
New England Journal of Medicine & 1275 & 1142 & 2245\\
The Lancet & 982 & 1596 & 2248\\
Cell & 956 & 1241 & 3060\\
PNAS & 929 & 1116 & 2926\\
Chemical Reviews & 905 & 84 & 263\\
JAMA & 818 & 634 & 1192\\
JACS & 759 & 543 & 1246\\
Physical Review Letters & 750 & 140 & 149\\
Nucleic Acids Research & 714 & 232 & 590\\
Circulation & 712 & 314 & 807\\
Advanced Materials & 709 & 170 & 464\\
Chemical Society Reviews & 709 & 0 & 0\\
Angewandte Chemie Int. Ed. & 697 & 340 & 522\\
Nature Genetics & 679 & 61 & 243\\
Nature Medicine & 676 & 494 & 863\\
Journal of Clinical Oncology & 666 & 381 & 600\\
Proceedings of the IEEE C.S. & 666 & n/a & n/a\\
Nature Communications & 634 & 567 & 1347\\
\bottomrule
\end{tabular}
\end{table}

The journal $z$-index was calculated in the same manner as the author $z$-index. Specifically, for each retracted paper, the journal impact factor IF in the year of publication was added to the journal's $z$-index. Similarly, the journal $\tilde{z}$-index was obtained by applying the temporal weighting factor based on the time elapsed between publication and retraction. Here, however, the solid curve in \cref{fig1} is used as $f(t)$, because journals do not voluntarily re-examine papers after publication, and thus no author-specific adjustment is needed. In addition, journal-specific delays are the responsibility of the journal itself. Given the current situation in which the retraction process often takes a long time, the function with $f(24)=2$ appears appropriate, as it would encourage journals to handle such processes more quickly. To perform these calculations, we used IF data made freely available by Clarivate~\cite{Clarivate}, covering the period from 1997 to 2025. As a result, retracted papers published before 1997 were not included in the calculations, and the resulting $z$- and $\tilde{z}$-indices should be regarded as conservative estimates.

\begin{figure}[h!t]
\centering
\includegraphics[width=1.0\linewidth]{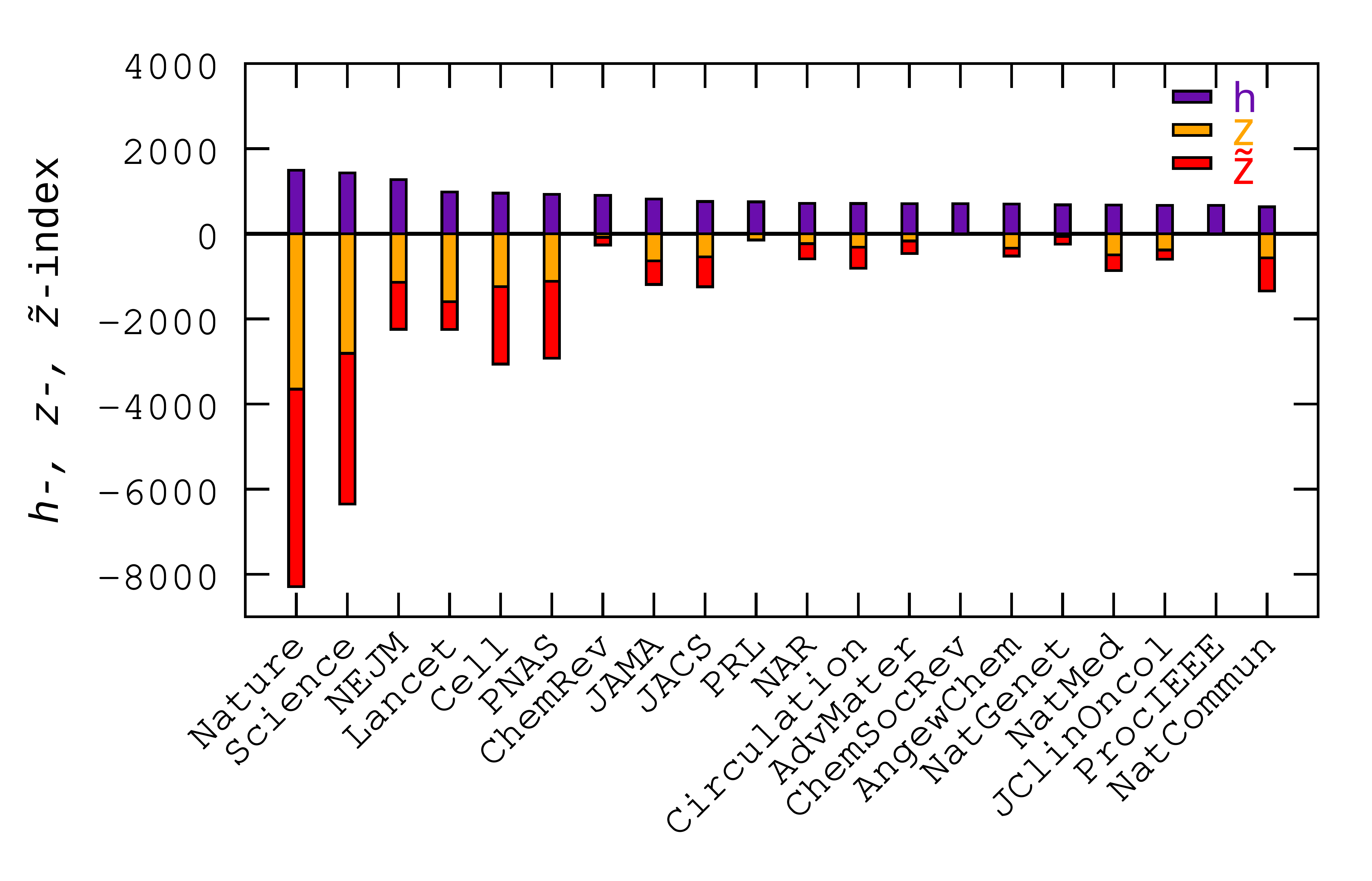}
\caption{Comparison of the $h$-, $z$-, and $\tilde{z}$-indices for the top 20 journals. Purple, orange, and red bars represent the $h$-, $z$-, and $\tilde{z}$-indices, respectively.}
\label{fig2}
\end{figure}

The statistical results are presented as a bar chart in \cref{fig2}. The purple bars represent the $h$-index, while the orange and red bars represent the $z$- and $\tilde{z}$-indices, respectively, stacked in the negative region of the $y$-axis. The journals are ordered according to their $h$-indices.

The most striking feature of \cref{fig2} is the extraordinarily large $\tilde{z}$-indices of {\it Nature} and {\it Science}. Their achievements, as reflected in their $h$-indices and accumulated over 157 and 146 years since their founding, respectively, appear to be outweighed by the negative achievements accumulated over only about 30 years, from 1997 to 2025. A grand-looking journal rests on a substantial amount of scientific debris. However, this should not be dismissed as the unavoidable fate of top journals. Excluding {\it Chemical Reviews}, {\it Chemical Society Reviews}, and {\it Proceedings of the IEEE}, which are not well suited to the present comparison because of their journal characteristics, the $h$-index exceeds the $\tilde{z}$-index in 6 of the remaining 15 journals. If the comparison is limited to the original $z$-index, the $h$-index exceeds the $z$-index in as many as 12 of the 15 journals, across a broad range of research fields.

The relatively low $z$-indices of {\it Physical Review Letters} and {\it Advanced Materials} may be explained by the possibility that retraction cases are less frequent in materials science than in the life sciences. {\it Nature} and {\it Science} by contrast include the life sciences within their scope. However, this explanation cannot account for the difference from the {\it New England Journal of Medicine (NEJM)}, which ranks third in the $h$-index. Its $\tilde{z}$-index is only about one-third to one-fourth that of {\it Nature} and {\it Science} despite having an $h$-index comparable to theirs.

The {\it NEJM} is a clinical medical journal and has a strong ethical commitment to the responsible sharing of data generated by clinical trials. Specifically, in 2017, the {\it NEJM} stated that it envisions a global research community in which sharing data becomes the norm~\cite{NEJMstatements}. Although it is unclear whether this statement has been effective, only six papers have been retracted in the nine years since then, and four of these were retracted within the short period of one month~\cite{NEJMret1,NEJMret2,NEJMret3,NEJMret4,NEJMret5,NEJMret6}. During the same period, {\it Nature} retracted 25 papers, with an average retraction time of 25.8 months. Thus, the trends in $\tilde{z}$ for these two journals move in opposite directions. Their policies on data sharing are also opposite: {\it Nature} recognizes the right of authors \emph{not} to share their data~\cite{GrafLetter}.

Another notable feature of \cref{fig2} is that the $\tilde{z}$-index is more than twice as large as the corresponding $z$-index. This is not limited to {\it Nature} and {\it Science}, but can also be readily seen for journals such as {\it Cell} and {\it PNAS}. Because the temporal weighting function used to calculate $\tilde{z}$ was defined such that $f(24)=2$, these results suggest that the average time required for retraction is well in excess of two years. As discussed above, retractions resulting from honest error would generally be expected to proceed relatively quickly~\cite{COPE2025} and therefore contribute only modestly to the $\tilde{z}$-index. The prolonged retraction times observed in these journals suggest that they may have attracted papers whose problems cannot be explained simply as honest error, and that the journals may have been too cautious about retracting such papers.
\section*{Discussion}
The $h$-index can be viewed as a measure of the dissemination of reliable knowledge, whereas the $\tilde{z}$-index can be viewed as a measure of the dissemination of false knowledge. That is, $h-\tilde{z}$ represents the net scientific value. Current scientific indicators are, in practice, positive-only. The introduction of a negative indicator is therefore highly significant. Scientists may sometimes disseminate incorrect information, and journals may sometimes amplify it. However, the aim is not to condemn individuals but to reduce the spread of error. Now that an objective negative indicator is available, whether it can be reduced depends on the actions of the scientific community. If the time required for retraction could be shortened to one year, then, because $f(12)=0.5$, the current $\tilde{z}$ values could be reduced to less than one quarter of their present levels (see \cref{fig2}). Consequently, the $h$-index would exceed the $\tilde{z}$-index for almost all journals. If future monitoring of $\tilde{z}$ shows that journals have shortened retraction times, $f(t)$ should be revised as appropriate, for example by adopting a more intuitive linear function connecting the origin and $f(12)=1$. The shape of $f(t)$ is open to discussion.

In addition to reducing retraction delays, avoiding the publication of problematic papers is a more fundamental way to reduce $\tilde{z}$. In these confusing times, publishing journals may have a responsibility to verify that at least the figures can be reproduced from the raw data. This can be done during peer review if journals require authors to submit raw data with the manuscript. Publishing those raw data alongside the manuscript would also make the work of {\it PubPeer}~\cite{PubPeer} and {\it For Better Science}~\cite{ForBetterScience} more effective, allowing misconduct to be detected earlier and thereby reducing the journal's own $\tilde{z}$-index.

For this purpose, the so-called ``source data'' made available by some journals~\cite{natdatapolicy} are of course of no value.{\if0 Journals need only require \emph{untouched} raw data. They should not be tinkered with casually, nor should journals openly tolerate such practice. Otherwise, both author and journal $z$-indices will continue to grow without bound.{\if0 Data must be treated with greater care and respect, just as the {\it NEJM} regards them as the burden borne by patients.\fi}{\if0 Our data, too, rest on citizens' costs.\fi}{\if0 Sharing them as they are should be our ethical responsibility.\fi}\fi}

Finally, it may be useful to close the discussion by comparing the data policies of two journals that showed distinctive $\tilde{z}$-index patterns (see \cref{tab:nn}). What emerges from the table is the possibility that the difference in the $\tilde{z}$-index reflects a deeper contrast: respect for the data themselves versus respect for authors' discretion in handling the data. How strongly this difference in data policy will affect the gap in $\tilde{z}$, and whether the gap will continue to widen without limit, is an interesting question.
\begin{table*}[h!b]
\centering
\caption{Comparison of Data Policies between {\it {\bf Nature}}~\cite{natdatapolicy} and {\it {\bf New England Journal of Medicine}}~\cite{NEJMstatements}}
\label{tab:nn}
\begin{tabular*}{\textwidth}{@{\extracolsep{\fill}}L{0.2\textwidth} L{0.4\textwidth} L{0.34\textwidth}@{}}
 & Nature ($\tilde{z}=8290$) & NEJM ($\tilde{z}=2245$)\\
\midrule
Year of policy statement & 2016 & 2017\\
Data availability statement & Mandatory & Mandatory\\
Data sharing & Not mandatory, because authors have the right not to share data.\textsuperscript\textdagger & Not mandatory, but it may influence editorial decisions.\textsuperscript\textdaggerdbl\\
May data be processed before sharing? & Yes; that is what ``source data'' means. & Not explicitly allowed.\\
Retracted papers published after the policy statement & 26 & 6\\
\bottomrule
\end{tabular*}
\addtabletext{\textsuperscript\textdagger See ref.~\citenum{GrafLetter}. \textsuperscript\textdaggerdbl See ref.~\citenum{NEJMstatements}.}
\end{table*}
\matmethods{Author and journal $z$- and $\tilde{z}$-indices are easily computed from only journal impact factors~\cite{Clarivate} and the Retraction Watch Data~\cite{RetractionWatchDB} using custom Python scripts.
\subsection*{Use of generative AI}
ChatGPT (GPT-5.5; OpenAI) was used during manuscript preparation to assist with English translation, wording refinement, and discussion related to the Python scripts. The Python scripts and all AI-assisted outputs were reviewed, fact-checked, and edited as needed by the author, who takes full responsibility for the final manuscript.
}
\showmatmethods{} 

\dataavail{The Python scripts used to calculate the $z$- and $\tilde{z}$-indices, together with the processed datasets, are available at: \href{https://github.com/zen137a0/z-index}{https://github.com/zen137a0/z-index}, which have been permanently archived in Zenodo with DOI:~10.5281/zenodo.20843889~\cite{zenzenodo}. Redistribution and use in source and binary forms, with or without modification, are permitted for research and educational purposes only.}

\acknow{I am grateful to Jorge Hirsch for his encouragement regarding these ideas. I also thank him for pointing out excessive aspects of the original draft, in particular that the temporal weighting function was too severe. In response to his comments, I applied the relaxed function $f_{A}(t)$ to authors. For journals, however, I retained the original function, because journals benefit greatly from the impact factor system, and with such benefits comes responsibility.}
\showacknow{} 


\begin{thebibliography}{99}
\bibitem{Hirsch2005}
  J. E. Hirsch, An index to quantify an individual's scientific research output. \href{https://doi.org/10.1073/pnas.0507655102}{{\it Proc. Natl. Acad. Sci. U.S.A.} {\bf 102}, 16569--16572 (2005)}.
\bibitem{Hirsch2007}
  J. E. Hirsch, Does the $h$ index have predictive power? \href{https://doi.org/10.1073/pnas.0707962104}{{\it Proc. Natl. Acad. Sci. U.S.A.} {\bf 104}, 19193--19198 (2007)}.
\bibitem{Ball2007}
  P. Ball, ``Achievement index climbs the ranks.'' Nature News. \url{https://doi.org/10.1038/448737a}. Posted 16 August 2007.
\bibitem{Acuna2012}
  D. E. Acuna, S. Allesina, K. P. Kording, Predicting scientific success. \href{https://doi.org/10.1038/489201a}{{\it Nature} {\bf 489}, 201--202 (2012)}.
\bibitem{Hirsch2020}
  G. Conroy, ``What's wrong with the h-index, according to its inventor?'' Nature News. \url{https://www.nature.com/nature-index/news/whats-wrong-with-the-h-index-according-to-its-inventor}. Posted 24 March 2020.
\bibitem{vanKampen2022}
  M. van Kampen, ``Anatomy of a Retraction 2 – Superconductive Fraud.'' For Better Science. \url{https://forbetterscience.com/2022/10/12/anatomy-of-a-retraction-2-superconductive-fraud/}. Posted 12 October 2022.
\bibitem{Hand2022}
  E. Hand, ``‘Something is seriously wrong’: Room-temperature superconductivity study retracted.'' Science News. \url{https://doi.org/10.1126/science.adf0548}. Posted 26 September 2022.
\bibitem{GaristoAPS2023}
  D. Garisto, ``Allegations of Scientific Misconduct Mount as Physicist Makes His Biggest Claim Yet.'' Physics News. \url{https://physics.aps.org/articles/v16/40}. Posted 09 March 2023.
\bibitem{vanKampen2023}
  M. van Kampen, ``Superconductive Fraud: The Sequel.'' For Better Science. \url{https://forbetterscience.com/2023/03/29/superconductive-fraud-the-sequel/}. Posted 29 March 2023.
\bibitem{NYTimes2024}
  T. Rosenbluth, ``Physicist Who Made Superconductor Claims Exits University of Rochester.'' The New York Times. \url{https://www.nytimes.com/2024/11/19/science/ranga-dias-rochester-superconductor.html}. Posted 19 November 2024.
\bibitem{RangaH}
  Google Scholar, ``Ranga Dias.'' \url{https://scholar.google.com/citations?user=HDoF5YEAAAAJ&hl=en&oi=ao} (accessed 25 July 2026).
\bibitem{RangaNat2020}
  Elliot Snider, Nathan Dasenbrock-Gammon, Raymond McBride, Mathew Debessai, Hiranya Vindana, Kevin Vencatasamy, Keith V. Lawler, Ashkan Salamat, Ranga P. Dias, RETRACTED ARTICLE: Room-temperature superconductivity in a carbonaceous sulfur hydride. \href{https://doi.org/10.1038/s41586-020-2801-z}{{\it Nature} {\bf 586}, 373--377 (2020)}.
\bibitem{RangaNat2023}
  Nathan Dasenbrock-Gammon, Elliot Snider, Raymond McBride, Hiranya Pasan, Dylan Durkee, Nugzari Khalvashi-Sutter, Sasanka Munasinghe, Sachith E. Dissanayake, Keith V. Lawler, Ashkan Salamat, Ranga P. Dias, RETRACTED ARTICLE: Evidence of near-ambient superconductivity in a N-doped lutetium hydride. \href{https://doi.org/10.1038/s41586-023-05742-0}{{\it Nature} {\bf 615}, 244--250 (2023)}.
\bibitem{RangaPRL2021-1}
  Elliot Snider, Nathan Dasenbrock-Gammon, Raymond McBride, Xiaoyu Wang, Noah Meyers, Keith V. Lawler, Eva Zurek, Ashkan Salamat, Ranga P. Dias, RETRACTED: Synthesis of Yttrium Superhydride Superconductor with a Transition Temperature up to 262 K by Catalytic Hydrogenation at High Pressures. \href{https://doi.org/10.1103/PhysRevLett.126.117003}{{\it Phys. Rev. Lett.} {\bf 126}, 117003 (2021)}.
\bibitem{RangaPRL2021-2}
  Dylan Durkee, Nathan Dasenbrock-Gammon, G. Alexander Smith, Elliot Snider, Dean Smith, Christian Childs, Simon A.J. Kimber, Keith V. Lawler, Ranga P. Dias, Ashkan Salamat, RETRACTED: Colossal Density-Driven Resistance Response in the Negative Charge Transfer Insulator \ce{MnS2}. \href{https://doi.org/10.1103/PhysRevLett.127.016401}{{\it Phys. Rev. Lett.} {\bf 127}, 016401 (2021)}.
\bibitem{RangaChemComm2022}
  G. Alexander Smith, Ines E. Collings, Elliot Snider, Dean Smith, Sylvain Petitgirard, Jesse S. Smith, Melanie White, Elyse Jones, Paul Ellison, Keith V. Lawler, Ranga P. Dias, Ashkan Salamat, Retracted Article: Carbon content drives high temperature superconductivity in a carbonaceous sulfur hydride below 100 GPa. \href{https://doi.org/10.1039/D2CC03170A}{{\it Chem. Commun.} {\bf 58}, 9064--9067 (2022)}.
\bibitem{MagdalenaSkipper}
  Nature Portfolio: About our journals: Nature: Nature's mission statement. \url{https://www.nature.com/nature-portfolio/about-journals/nature} (accessed 25 July 2026).
\bibitem{Garfield2006}
  E. Garfield, The History and Meaning of the Journal Impact Factor. \href{https://doi.org/10.1001/jama.295.1.90}{{\it JAMA: Journal of the American Medical Association} {\bf 295}, 90--93 (2006)}; \url{https://garfield.library.upenn.edu/papers/jamajif2006.pdf}.
\bibitem{Clarivate}
  Clarivate Journal Citation Reports. \url{https://jcr.clarivate.com/jcr/home} (accessed 25 July 2026).
\bibitem{vanRaan2006}
  A. F. J. van Raan, Comparison of the Hirsch-index with standard bibliometric indicators and with peer judgment for 147 chemistry research groups. \href{https://doi.org/10.1556/Scient.67.2006.3.10}{{\it Scientometrics} {\bf 67}, 491--502 (2006)}.
\bibitem{Egghe2008}
  L. Egghe, Mathematical theory of the h- and g-index in case of fractional counting of authorship. \href{https://doi.org/10.1002/asi.20845}{{\it J. Am. Soc. Inf. Sci. Technol.} {\bf 59}, 1608--1616 (2008)}.
\bibitem{Schreiber2009}
  M. Schreiber, A case study of the modified Hirsch index $h_{m}$ accounting for multiple coauthors. \href{https://doi.org/10.1002/asi.21057}{{\it J. Am. Soc. Inf. Sci. Technol.} {\bf 60}, 1274--1282 (2009)}.
\bibitem{Hirsch2010hbar}
  J. E. Hirsch, An index to quantify an individual's scientific research output that takes into account the effect of multiple coauthorship. \href{https://doi.org/10.1007/s11192-010-0193-9}{{\it Scientometrics} {\bf 85}, 741--754 (2010)}.
\bibitem{Hirsch2019}
  J. E. Hirsch, $h_{\alpha}$: An index to quantify an individual's scientific leadership. \href{https://doi.org/10.1007/s11192-018-2994-1}{{\it Scientometrics} {\bf 118}, 673--686 (2019)}.
\bibitem{Tietze2020}
  A. Tietze, S. Galam, P. Hofmann, Crediting multi-authored papers to single authors. \href{https://doi.org/10.1016/j.physa.2020.124652}{{\it Physica A} {\bf 554}, 124652 (2020)}.
\bibitem{Tzitzikas2024}
  Y. Tzitzikas, G. Dovas, How co-authorship affects the H-index? \href{https://doi.org/10.1007/s11192-024-05088-y}{{\it Scientometrics} {\bf 129}, 4437--4469 (2024)}.
\bibitem{ICMJE-authorship}
  International Committee of Medical Journal Editors, ``Defining the Role of Authors and Contributors.'' \url{https://www.icmje.org/recommendations/browse/roles-and-responsibilities/defining-the-role-of-authors-and-contributors.html} (accessed 25 July 2026).
\bibitem{RetNoteRangaNat2020}
  The editors of {\it Nature}, Retraction Note: Room-temperature superconductivity in a carbonaceous sulfur hydride. \href{https://doi.org/10.1038/s41586-022-05294-9}{{\it Nature} {\bf 610}, 804 (2022)}.
\bibitem{Marel-Hirsch2022}
  D. van der Marel, J. E. Hirsch, Comment on {\it Nature} 586, 373 (2020) by E. Snider~{\it et al.} \href{https://arxiv.org/abs/2201.07686v1}{arXiv:2201.07686v1} (submission 19 January 2022).
\bibitem{Marel-Hirsch2023}
  D. van der Marel, J. E. Hirsch, Room-temperature superconductivity — or not? Comment on {\it Nature} 586, 373 (2020) by E. Snider~{\it et al.} \href{https://doi.org/10.1142/S0217979223750012}{{\it Int. J. Mod. Phys. B} {\bf 37}, 2375001 (2023)}.
\bibitem{SemenzaH}
  ScholarGPS, ``Gregg L. Semenza.'' \url{https://scholargps.com/scholars/38652215758338/gregg-l-semenza} (accessed 25 July 2026).
\bibitem{SudhofH}
  Google Scholar, ``Thomas C. Sudhof.'' \url{https://scholar.google.com/citations?user=-Z1g0SwAAAAJ&hl=en} (accessed 25 July 2026).
\bibitem{Schneider2020-sem}
  L. Schneider, ``Gregg Semenza: real Nobel Prize and unreal research data.'' For Better Science. \url{https://forbetterscience.com/2020/10/07/gregg-semenza-real-nobel-prize-and-unreal-research-data/}. Posted 07 October 2020.
\bibitem{vanKampen2023-sud}
  M. van Kampen, ``Thomas Südhof and the standards of scientific rigor.'' For Better Science. \url{https://forbetterscience.com/2023/10/04/thomas-sudhof-and-the-standards-of-scientific-rigor/}. Posted 04 October 2023.
\bibitem{PubPeer-sem}
  PubPeer database, ``Gregg Semenza.'' \url{https://pubpeer.com/search?q=Gregg+Semenza} (accessed 25 July 2026).
\bibitem{PubPeer-sud}
  PubPeer database, ``Thomas Sudhof.'' \url{https://pubpeer.com/search?q=Thomas+Sudhof} (accessed 25 July 2026).
\bibitem{sem1}
  Kenji Kasuno, Satoshi Takabuchi, Kazuhiko Fukuda, Shinae Kizaka-Kondoh, Junji Yodoi, Takehiko Adachi, Gregg L. Semenza, Kiichi Hirota, Retraction: Nitric Oxide Induces Hypoxia-inducible Factor 1 Activation That Is Dependent on MAPK and Phosphatidylinositol 3-Kinase Signaling. \href{https://doi.org/10.1074/jbc.M308197200}{{\it J. Biol. Chem.} {\bf 279}, 2550--2558 (2004)}.
\bibitem{sem2}
  Kiichi Hirota, Ryo Fukuda, Satoshi Takabuchi, Shinae Kizaka-Kondoh, Takehiko Adachi, Kazuhiko Fukuda, Gregg L. Semenza, Retraction: Induction of Hypoxia-inducible Factor 1 Activity by Muscarinic Acetylcholine Receptor Signaling. \href{https://doi.org/10.1074/jbc.M405164200}{{\it J. Biol. Chem.} {\bf 279}, 41521--41528 (2004)}.
\bibitem{sem3}
  Mariko Tomita, Gregg L. Semenza, Canine Michiels, Takehiro Matsuda, Jun-Nosuke Uchihara, Taeko Okudaira, Yuetsu Tanaka, Naoya Taira, Kazuiku Ohshiro, Naoki Mori, Retraction: Activation of hypoxia-inducible factor 1 in human T-cell leukaemia virus type 1-infected cell lines and primary adult T-cell leukaemia cells. \href{https://doi.org/10.1042/BJ20070286}{{\it Biochem. J.} {\bf 406}, 317--323 (2007)}.
\bibitem{sem4}
  KangAe Lee, David Z. Qian, Sergio Rey, Hong Wei, Jun O. Liu, Gregg L. Semenza, RETRACTED: Anthracycline chemotherapy inhibits HIF-1 transcriptional activity and tumor-induced mobilization of circulating angiogenic cells. \href{https://doi.org/10.1073/pnas.0812801106}{{\it Proc. Natl. Acad. Sci. U.S.A.} {\bf 106}, 2353--2358 (2009)}.
\bibitem{sem5}
  KangAe Lee, Huafeng Zhang, David Z. Qian, Sergio Rey, Jun O. Liu, Gregg L. Semenza, RETRACTED: Acriflavine inhibits HIF-1 dimerization, tumor growth, and vascularization. \href{https://doi.org/10.1073/pnas.0909353106}{{\it Proc. Natl. Acad. Sci. U.S.A.} {\bf 106}, 17910--17915 (2009)}.
\bibitem{sem6}
  H. Zhang, C. C. L. Wong, H. Wei, D. M. Gilkes, P. Korangath, P. Chaturvedi, L. Schito, J. Chen, B. Krishnamachary, P. T. Winnard Jr., V. Raman, L. Zhen, W. A. Mitzner, S. Sukumar, G. L. Semenza, RETRACTED ARTICLE: HIF-1-dependent expression of angiopoietin-like 4 and L1CAM mediates vascular metastasis of hypoxic breast cancer cells to the lungs. \href{https://doi.org/10.1038/onc.2011.365}{{\it Oncogene} {\bf 31}, 1757--1770 (2012)}.
\bibitem{sem7}
  Hong Wei, Djahida Bedja, Norimichi Koitabashi, Dongmei Xing, Jasper Chen, Karen Fox-Talbot, Rosanne Rouf, Shaoping Chen, Charles Steenbergen, John W. Harmon, Harry C. Dietz, Kathleen L. Gabrielson, David A. Kass, Gregg L. Semenza, RETRACTED: Endothelial expression of hypoxia-inducible factor 1 protects the murine heart and aorta from pressure overload by suppression of TGF-$\beta$ signaling. \href{https://doi.org/10.1073/pnas.1202081109}{{\it Proc. Natl. Acad. Sci. U.S.A.} {\bf 109}, E841--E850 (2012)}.
\bibitem{sem8}
  Daniele M. Gilkes, Saumendra Bajpai, Pallavi Chaturvedi, Denis Wirtz, Gregg L. Semenza, Withdrawal: Hypoxia-inducible Factor 1 (HIF-1) Promotes Extracellular Matrix Remodeling under Hypoxic Conditions by Inducing P4HA1, P4HA2, and PLOD2 Expression in Fibroblasts. \href{https://doi.org/10.1074/jbc.M112.442939}{{\it J. Biol. Chem.} {\bf 288}, 10819--10829 (2013)}.
\bibitem{sem9}
  Guoxiang Yuan, Ying-Jie Peng, Vaddi Damodara Reddy, Vladislav V. Makarenko, Jayasri Nanduri, Shakil A. Khan, Joseph A. Garcia, Ganesh K. Kumar, Gregg L. Semenza, Nanduri R. Prabhakar, RETRACTED: Mutual antagonism between hypoxia-inducible factors 1$\alpha$ and 2$\alpha$ regulates oxygen sensing and cardio-respiratory homeostasis. \href{https://doi.org/10.1073/pnas.1305961110}{{\it Proc. Natl. Acad. Sci. U.S.A.} {\bf 110}, E1788--E1796 (2013)}.
\bibitem{sem10}
  Daniele M. Gilkes, Saumendra Bajpai, Carmen C. Wong, Pallavi Chaturvedi, Maimon E. Hubbi, Denis Wirtz, Gregg L. Semenza, Retraction: Procollagen lysyl hydroxylase 2 is essential for hypoxia-induced breast cancer metastasis. \href{https://doi.org/10.1158/1541-7786.MCR-12-0629}{{\it Mol. Cancer Res.} {\bf 11}, 456--466 (2013)}.
\bibitem{sem11}
  Daniele M. Gilkes, Pallavi Chaturvedi, Saumendra Bajpai, Carmen C. Wong, Hong Wei, Stephen Pitcairn, Maimon E. Hubbi, Denis Wirtz, Gregg L. Semenza, RETRACTED: Collagen prolyl hydroxylases are essential for breast cancer metastasis. \href{https://doi.org/10.1158/0008-5472.CAN-12-3963}{{\it Cancer Res.} {\bf 73}, 3285--3296 (2013)}.
\bibitem{sem12}
  Daniele M. Gilkes, Lisha Xiang, Sun Joo Lee, Pallavi Chaturvedi, Maimon E. Hubbi, Denis Wirtz, Gregg L. Semenza, RETRACTED: Hypoxia-inducible factors mediate coordinated RhoA-ROCK1 expression and signaling in breast cancer cells. \href{https://doi.org/10.1073/pnas.1321510111}{{\it Proc. Natl. Acad. Sci. U.S.A.} {\bf 111}, E384--E393 (2013)}.
\bibitem{sem13}
  Debangshu Samanta, Daniele M. Gilkes, Pallavi Chaturvedi, Lisha Xiang, Gregg L. Semenza, RETRACTED: Hypoxia-inducible factors are required for chemotherapy resistance of breast cancer stem cells. \href{https://doi.org/10.1073/pnas.1421438111}{{\it Proc. Natl. Acad. Sci. U.S.A.} {\bf 111}, E5429--E5438 (2014)}.
\bibitem{sem14}
  Debangshu Samanta, Youngrok Park, Shaida A. Andrabi, Laura M. Shelton, Daniele M. Gilkes, Gregg L. Semenza, RETRACTED: PHGDH expression is required for mitochondrial redox homeostasis, breast cancer stem cell maintenance, and lung metastasis. \href{https://doi.org/10.1158/0008-5472.CAN-16-0530}{{\it Cancer Res.} {\bf 76}, 4430--4442 (2016)}.
\bibitem{sem15}
  Huafeng Zhang, Marta Bosch-Marce, Larissa A. Shimoda, Yee Sun Tan, Jin Hyen Baek, Jacob B. Wesley, Frank J. Gonzalez, Gregg L. Semenza, Withdrawal: Mitochondrial Autophagy Is an HIF-1-dependent Adaptive Metabolic Response to Hypoxia. \href{https://doi.org/10.1074/jbc.M800102200}{{\it J. Biol. Chem.} {\bf 283}, 10892--10903 (2008)}.
\bibitem{sud1}
  Lulu Y. Chen, Man Jiang, Bo Zhang, Ozgun Gokce, Thomas C. S\"{u}dhof, RETRACTED: Conditional Deletion of All Neurexins Defines Diversity of Essential Synaptic Organizer Functions for Neurexins. \href{https://doi.org/10.1016/j.neuron.2017.04.011}{{\it Neuron} {\bf 94}, 611--625 (2017)}.
\bibitem{sud2}
  Pei-Yi Lin, Lulu Y. Chen, Peng Zhou, Sung-Jin Lee, Justin H. Trotter, Thomas C. S\"{u}dhof, RETRACTED: Neurexin-2 restricts synapse numbers and restrains the presynaptic release probability by an alternative splicing-dependent mechanism. \href{https://doi.org/10.1073/pnas.2300363120}{{\it Proc. Natl. Acad. Sci. U.S.A.} {\bf 120}, e2300363120 (2023)}.
\bibitem{Fang2012}
  F. C. Fang, R. G. Steen, A. Casadevall, Misconduct accounts for the majority of retracted scientific publications. \href{https://doi.org/10.1073/pnas.1212247109}{{\it Proc. Natl. Acad. Sci. U.S.A.} {\bf 109}, 17028--17033 (2012)}.
\bibitem{COPE2025}
  COPE Council, ``COPE Retraction guidelines,'' section ``Importance of timely handling of retractions.'' \url{https://www.calismatoplum.org/wp-content/uploads/2025/09/retraction-guidelines-cope-1.pdf} (VERSION 3: August 2025).
\bibitem{Budd1998}
  J. M. Budd, M.E. Sievert, T. R. Schultz, Phenomena of Retraction---Reasons for Retraction and Citations to the Publications. \href{https://doi.org/10.1001/jama.280.3.296}{{\it J. Am. Med. Assoc.} {\bf 280}, 296--297 (1998)}.
\bibitem{BarIlan2018}
  J. Bar-Ilan, G. Halevi, Temporal characteristics of retracted articles. \href{https://doi.org/10.1007/s11192-018-2802-y}{{\it Scientometrics} {\bf 116}, 1771--1783 (2018)}.
\bibitem{Candal2020}
  C. Candal-Pedreira, A. Ruano-Ravina, E. Fernández, J. Ramos, I. Campos-Varela,  M. Pérez-Ríos, Does retraction after misconduct have an impact on citations? A pre-post study. \href{https://doi.org/10.1136/bmjgh-2020-003719}{{\it BMJ Glob. Health} {\bf 5}, e003719 (2020)}.
\bibitem{Hsiao2022}
  T.-K. Hsiao, J. Schneider, Continued use of retracted papers: Temporal trends in citations and (lack of) awareness of retractions shown in citation contexts in biomedicine. \href{https://doi.org/10.1162/qss_a_00155}{{\it Quant. Sci. Stud.} {\bf 2}, 1144--1169 (2022)}.
\bibitem{Garfield2003}
  E. Garfield, The meaning of the Impact Factor. \href{https://www.aepc.es/ijchp/articulos.php?coid=English&id=77}{{\it Int. J. Clin. Health Psychol.} {\bf 3}, 363--369 (2003)}.
\bibitem{vanderVet2016}
  P. E. van der Vet, H. Nijveen, Propagation of errors in citation networks: a study involving the entire citation network of a widely cited paper published in, and later retracted from, the journal Nature. \href{https://doi.org/10.1186/s41073-016-0008-5}{{\it Res. Integr. Peer Rev.} {\bf 1}, 3 (2016)}.
\bibitem{Oransky2025}
  I. Oransky, A. Marcus, Retraction Reactions. \href{https://doi.org/10.1511/2025.113.4.208}{{\it Am. Sci.} {\bf 113}, 208 (2025)}.
\bibitem{Cho2019}
  Inha Cho, Zhi-Jun Jia, Frances H. Arnold, RETRACTED: Site-selective enzymatic C‒H amidation for synthesis of diverse lactams. \href{https://doi.org/10.1126/science.aaw9068}{{\it Science} {\bf 364}, 575--578 (2019)}.
\bibitem{ArnoldH}
  Google Scholar, ``Frances Arnold.'' \url{https://scholar.google.com/citations?user=wil5NhcAAAAJ&hl=en} (accessed 25 July 2026).
\bibitem{SCImago}
  SCImago Rankings. \url{https://www.scimagojr.com/journalrank.php?order=h&ord=desc&page=1&total_size=32193} (accessed 25 July 2026).
{}
\bibitem{NEJMstatements}
  D. B. Taichman, P. Sahni, A. Pinborg, L. Peiperl, C. Laine, A. James, S.-T. Hong, A. Haileamlak, L. Gollogly, F. Godlee, F. A. Frizelle, F. Florenzano, J. M. Drazen, H. Bauchner, C. Baethge, J. Backus, Data Sharing Statements for Clinical Trials---A Requirement of the International Committee of Medical Journal Editors. \href{https://doi.org/10.1056/NEJMe1705439}{{\it N. Engl. J. Med.} {\bf 376}, 2277--2279 (2017)}.
\bibitem{NEJMret1}
  José R. Banegas, Luis M. Ruilope, Alejandro de la Sierra, Ernest Vinyoles, Manuel Gorostidi, Juan J. de la Cruz, Gema Ruiz-Hurtado, Julián Segura, Fernando Rodríguez-Artalejo, Bryan Williams, RETRACTED: Relationship between Clinic and Ambulatory Blood-Pressure Measurements and Mortality. \href{https://doi.org/10.1056/NEJMoa1712231}{{\it N. Engl. J. Med.} {\bf 378}, 1509--1520 (2018)}.
\bibitem{NEJMret2}
  Marion H. Brown, Michael L. Dustin, RETRACTED: Steering CAR T Cells into Solid Tumors. \href{https://doi.org/10.1056/NEJMcibr1811991}{{\it N. Engl. J. Med.} {\bf 380}, 289--291 (2019)}.
\bibitem{NEJMret3}
  Jitender Jinagal, Poonam Dhiman, RETRACTED: Retinal Hemorrhage from Blunt Ocular Trauma. \href{https://doi.org/10.1056/NEJMicm1905974}{{\it N. Engl. J. Med.} {\bf 381}, 2252 (2019)}.
\bibitem{NEJMret4}
  Mandeep R. Mehra, Sapan S. Desai, SreyRam Kuy, Timothy D. Henry, Amit N. Patel, RETRACTED: Cardiovascular Disease, Drug Therapy, and Mortality in Covid-19. \href{https://doi.org/10.1056/NEJMoa2007621}{{\it N. Engl. J. Med.} {\bf 382}, e102 (2020)}.
\bibitem{NEJMret5}
  David R.W. Jayne,  Peter A. Merkel, Thomas J. Schall, Pirow Bekker, RETRACTED: Avacopan for the Treatment of ANCA-Associated Vasculitis. \href{https://doi.org/10.1056/NEJMoa2023386}{{\it N. Engl. J. Med.} {\bf 384}, 599--609 (2021)}.
\bibitem{NEJMret6}
  Yuling Wang, Xiangdong Mu, RETRACTED: Bronchial Casts from Inhalation of Forest-Fire Smoke. \href{https://doi.org/10.1056/NEJMicm2518379}{{\it N. Engl. J. Med.} {\bf 394}, 1634 (2026)}.
\bibitem{GrafLetter}
  Letter from the Research Integrity Director at Springer Nature. \url{https://forbetterscience.com/wp-content/uploads/2024/05/chrisgrafletter.pdf}. Dated 29 September 2023.
\bibitem{PubPeer}
  PubPeer. \url{https://pubpeer.com/}.
\bibitem{ForBetterScience}
  For Better Science. \url{https://forbetterscience.com/}.
\bibitem{natdatapolicy}
  Nature: For Authors: Initial submission. See the link ``data availability and data citations policy,'' under the section ``Data availability.'' \url{https://www.nature.com/nature/for-authors/initial-submission}. Dated: September 2016.
\bibitem{RetractionWatchDB}
  Retraction-watch-data. \url{https://gitlab.com/crossref/retraction-watch-data} (accessed 25 July 2026).
\bibitem{zenzenodo}
  N. Zen, zen137a0/z-index: v2.3. Zenodo. \href{https://doi.org/10.5281/zenodo.20843889}{https://doi.org/10.5281/zenodo.20843889}. Deposited 01 July 2026.
\end{thebibliography}
\urlstyle{same}
\balance

\end{document}